\documentclass[twocolumn,prl,9pt,superscriptaddress,amsfonts,nobibnotes]{revtex4-1}

\input{glyphtounicode}
\usepackage[T1]{fontenc}
\usepackage[utf8]{inputenc}
\usepackage[pdftex]{graphicx}
\usepackage{xcolor}
\usepackage[pdftex,bookmarks=true,bookmarksopen,bookmarksnumbered,
colorlinks,
linkcolor=blue,
citecolor=blue
]{hyperref}
\usepackage{bm}
\usepackage{newfloat}
\usepackage{physics}
\usepackage{siunitx}
\usepackage{multibib}
\usepackage{textcomp}

\newcites{method}{Methods References}

\makeatletter
\renewcommand{\@caption@fignum@sep}{{\bfseries~|~}}
\renewcommand{\figurename}{Figure}
\renewcommand{\tablename}{Table}
\renewcommand{\fnum@figure}{{\boldmath\bfseries\figurename~\thefigure}}
\renewcommand{\fnum@table}{{\boldmath\bfseries\tablename~\thetable}}
\DeclareFloatingEnvironment[
    fileext=los,
    name=Figure,
    placement=tbhp,
    within=none,
]{extfig}
\renewcommand{\extfigname}{Extended Data Figure}
\renewcommand{\fnum@extfig}{{\normalfont\bfseries\extfigname~\theextfig}}
\makeatother

\let\Oldvphantom\vphantom
\renewcommand{\vphantom}[1]{\mathinner{\!\Oldvphantom{#1}}}

\pdfpageattr {/Group << /S /Transparency /I true /CS /DeviceRGB>>} 

\begin{document}
	
	\title{Bell-state tomography in a silicon many-electron artificial molecule}
	\makeatletter
	\author{Ross C. C. Leon}
	\email[r.leon@unsw.edu.au]{}
	\affiliation{School of Electrical Engineering and Telecommunications, The University of New South Wales, Sydney, NSW 2052, Australia.}
	\author{Chih Hwan Yang}
	\affiliation{School of Electrical Engineering and Telecommunications, The University of New South Wales, Sydney, NSW 2052, Australia.}
	\author{Jason C. C. Hwang}	
	\altaffiliation{Current address: Research and Prototype Foundry, The University of Sydney, Sydney, NSW 2006, Australia.}
	\affiliation{School of Electrical Engineering and Telecommunications, The University of New South Wales, Sydney, NSW 2052, Australia.}
	\author{Julien Camirand Lemyre}
	\affiliation{Institut Quantique et D\'epartement de Physique, Universit\'e de Sherbrooke, Sherbrooke, Qu\'ebec J1K 2R1, Canada}
	\author{Tuomo Tanttu}
	\affiliation{School of Electrical Engineering and Telecommunications, The University of New South Wales, Sydney, NSW 2052, Australia.}
	\author{Wei Huang}
	\affiliation{School of Electrical Engineering and Telecommunications, The University of New South Wales, Sydney, NSW 2052, Australia.}
	\author{Jonathan Y. Huang}
	\affiliation{School of Electrical Engineering and Telecommunications, The University of New South Wales, Sydney, NSW 2052, Australia.}
	\author{Fay E. Hudson}
	\affiliation{School of Electrical Engineering and Telecommunications, The University of New South Wales, Sydney, NSW 2052, Australia.}
	\author{Kohei M. Itoh}
	\affiliation{School of Fundamental Science and Technology, Keio University, 3-14-1 Hiyoshi, Kohokuku, Yokohama 223-8522, Japan.}
	\author{Arne Laucht}
	\affiliation{School of Electrical Engineering and Telecommunications, The University of New South Wales, Sydney, NSW 2052, Australia.}
	\author{Michel Pioro-Ladri\`ere}
	\affiliation{Institut Quantique et D\'epartement de Physique, Universit\'e de Sherbrooke, Sherbrooke, Qu\'ebec J1K 2R1, Canada}
	\affiliation{Quantum Information Science Program, Canadian Institute for Advanced Research, Toronto, ON, M5G 1Z8, Canada}
	\author{Andre Saraiva}
	\email[a.saraiva@unsw.edu.au]{}
	\affiliation{School of Electrical Engineering and Telecommunications, The University of New South Wales, Sydney, NSW 2052, Australia.}
	\author{Andrew S. Dzurak}
	\email[a.dzurak@unsw.edu.au]{}
	\affiliation{School of Electrical Engineering and Telecommunications, The University of New South Wales, Sydney, NSW 2052, Australia.}
	\makeatother

	\maketitle

\phantomsection
\addcontentsline{toc}{section}{ Introduction}
{\bfseries\boldmath{
An error-corrected quantum processor will require millions of qubits~\cite{campbell_roads_2017}, accentuating the advantage of nanoscale devices with small footprints, such as silicon quantum dots~\cite{zwanenburg_silicon_2013}. However, as for every device with nanoscale dimensions, disorder at the atomic level is detrimental to qubit uniformity.
Here we investigate two spin qubits confined in a silicon double-quantum-dot artificial molecule. Each quantum dot has a robust shell structure and, when operated at an occupancy of 5 or 13 electrons, has single spin-\textonehalf\ valence electron in its \textit{p}- or \textit{d}-orbital, respectively~\cite{leon2020coherent}. 
These higher electron occupancies screen atomic-level disorder~\cite{barnes2011screening,hu2001spinbased,leon2020coherent}.
The larger multielectron wavefunctions also enable significant overlap between neighbouring qubit electrons, while making space for an interstitial exchange-gate electrode.
We implement a universal gate set using the magnetic field gradient of a micromagnet for electrically-driven 
single qubit gates~\cite{pioro-ladriere2008electrically}, and a gate-voltage-controlled inter-dot barrier to perform two-qubit gates by pulsed exchange coupling. We use this gate set to demonstrate a Bell state preparation between multielectron qubits with fidelity \SI{90.3}{\%}, confirmed by two-qubit state tomography using spin parity measurements~\cite{seedhouse2020parity}.
}}
	
\phantomsection
\addcontentsline{toc}{section}{ Main}

Semiconductor nanodevices, especially those incorporating oxide insulating layers, suffer from variability due to various atomic-scale defects and morphological imprecision. This disorder degrades spin qubit performance due to the sub-nanometre wave properties of single electrons. The conflict between the benefits of densely packing many quantum dots within a chip and the exposure to disorder demands further research regarding improved systems for encoding solid-state qubits. We exploit here the operation of qubits in silicon metal-oxide-semiconductor (Si-MOS) quantum dots containing several electrons that form closed shells, leaving a single valence electron in the outer shell~\cite{leon2020coherent}. The spin of a valence electron in a high-occupancy Si-MOS quantum dot was previously shown to form a high-fidelity single qubit~\cite{leon2020coherent}, at least in part due to the improved screening of disorder provided by the raised electron density. However, it was not clear how well two-qubit logic could be performed using such systems, because of the complex molecular states present in a many-electron double quantum dot~\cite{hu2001spinbased}. We address this here using two multielectron qubits to operate an isolated quantum processing unit~\cite{watson2018programmable,yang2020operation}.

\begin{figure*}
	\centering
	\includegraphics[width=\linewidth]{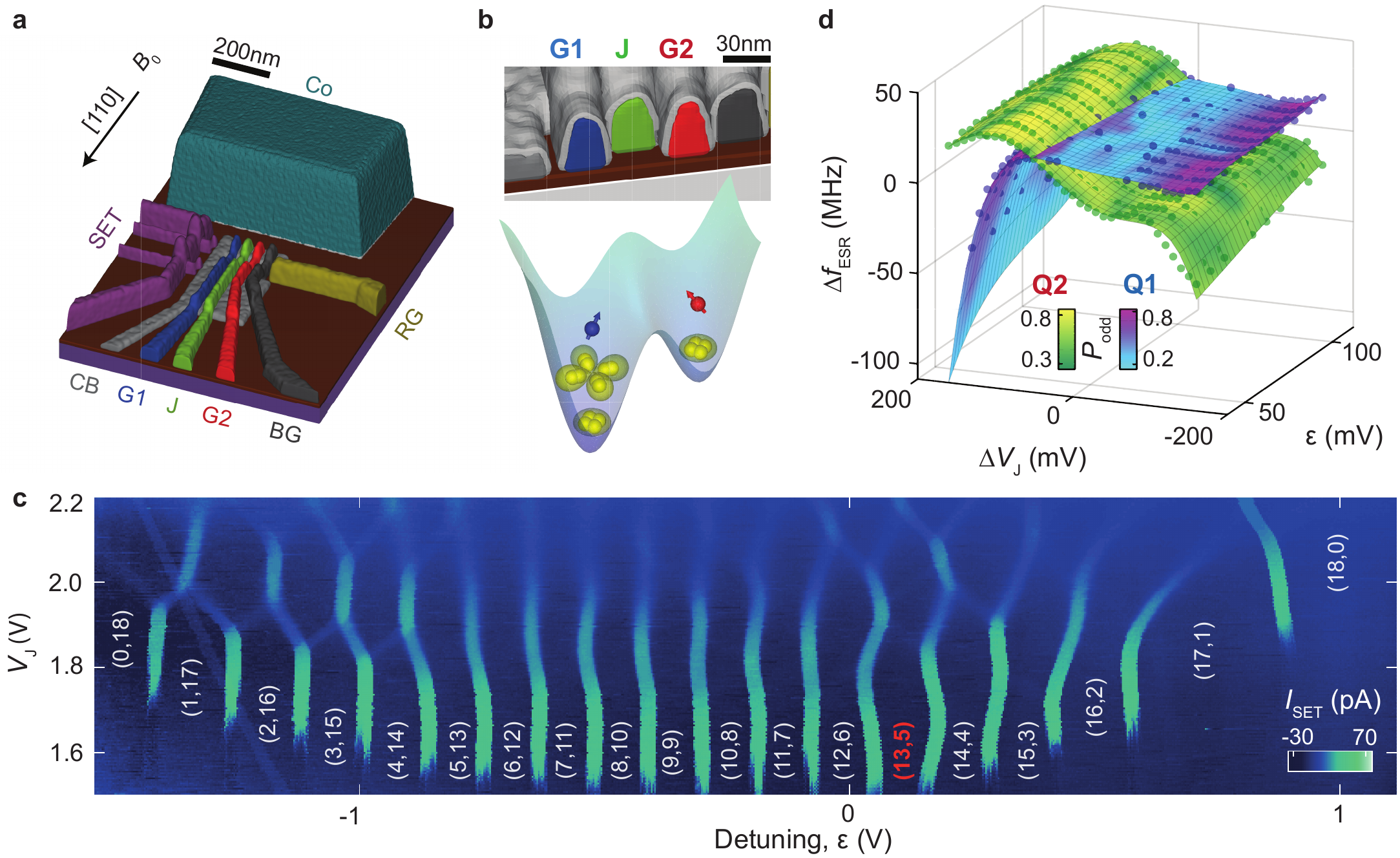}
	\caption{\textbf{Device overview and electron occupancy measurement.} 
		\textbf{a}, A 3D visualisation of the Si-MOS device structure. A quantum dot is formed under gate G1 (blue) and G2 (red), with inter-dot tunnel rates controlled by J (green). Gate RG enables connection to an n-doped reservoir to load/unload electrons to/from the quantum dot, with tunnel rates controlled by BG. Gate CB serves as a confinement barrier in lateral direction. The cobalt structure at the top of the image acts as both a micromagnet and electrode for EDSR control (dark green).
		\textbf{b}, Top: Cross-section diagram of panel \textbf{(a)} along the $[1\Bar{1}0]$ crystallographic direction. Bottom: Schematic showing the number of electrons in each of the two quantum dots, aligning with the metal gates in the panel above. The height of each electron represents its relative energy and the shell to which it belongs. Yellow electrons form full shells and are inert, while the extra electron in each dot (blue and red) act as an effective single spin qubit.
		\textbf{c}, Charge stability map of the double quantum dot at $B_0=\SI{0}{\tesla}$, showing the charge occupancies ($N_1$,$N_2$), produced by plotting the lock-in signal from SET sensor $I_\mathrm{SET}$ as a function of detuning $\varepsilon$ and $V_\mathrm{J}$. The detuning $\varepsilon =  V_\mathrm{G1}-V_\mathrm{G2}$ is referenced by $\varepsilon=\SI{0}{\V}$ at the charge readout transition (12,6)$\Longleftrightarrow$(13,5). A square wave with peak-to-peak amplitude of \SI{2}{\mV} and frequency \SI{487}{\Hz} is applied to G1 for lock-in excitation. Dynamic compensation is applied to the SET sensor to maintain a high readout sensitivity.  
		\textbf{d}, Resonance frequency of Q1 and Q2 as a function of $\varepsilon$ and $V_\mathrm{J}$ with $\ket{\downarrow\downarrow}$ initialisation. Color scale represents the adiabatic inversion probability.
	}\label{SEMs}
\end{figure*}
	
	\bigskip

This demonstration is performed with the device structure depicted in \autoref{SEMs}a, and investigated in previous studies~\cite{yang2020operation,leon2020coherent}. Using the technique adopted from Ref.~\onlinecite{yang2020operation}, where the quantum dots are isolated from the electron reservoir, we load electrons into the two quantum dots formed under gates G1 and G2 and separated by gate J. We monitor inter-dot charge transitions by measuring the transconductance of a nearby single electron transistor (SET). An on-chip cobalt micromagnet is fabricated \SI{120}{\nm} away from the quantum dots. This micromagnet serves two purposes: to create an inhomogeneous magnetic field as well as an oscillatory electric field, for electrically-driven spin resonance (EDSR)~\cite{pioro-ladriere2008electrically,takeda2016faulttolerant,zajac2018resonantly}.
	
In order to achieve an isolated mode of operation, the quantum dots are initialised with a desired number of electrons using the reservoir under RG, then the tunnel rate between the quantum dot under G2 and the reservoir is made negligible by lowering the voltage applied to gate BG, such that the double quantum dot becomes isolated~\cite{yang2020operation}. \autoref{SEMs}c is a charge stability diagram with vertical lines indicating inter-dot charge transition. For the experiment discussed here, we load a total of 18 electrons. Note that diagonal lines on the upper half of \autoref{SEMs}c (around $V_\mathrm{J}$ = \SI{1.9}{\V}) mark transitions in which the J gate becomes too attractive for electrons, and instead of forming a barrier it forms a quantum dot between G1 and G2~\cite{yang2020operation}. At very low voltages, the J gate creates a large barrier between the dots suppressing inter-dot tunnelling. Once the tunnel rate becomes lesser than the lock-in frequency (\SI{487}{\Hz}), the transition lines fade, as observed for $V_\mathrm{J} < \SI{1.6}{\V}$.

In a small two dimensional circular quantum dot, full shells are formed at 4 and 12 electrons \cite{yang2013spinvalley,leon2020coherent, camenzind_spectroscopy_2019,liles_spin_2018}. The fourfold degeneracy of the first shell has its origin in the spin and valley degrees of freedom for silicon conduction band electrons. 
The next shell is formed by two-dimensional p-like states, which means the $p_x$ and $p_y$ states are quasi-degenerate in the approximately circularly symmetric dot. This shell can fit a total of 8 electrons. 
We control the voltage detuning $\varepsilon$ between gates G1 and G2 voltages such that there are 13 and 5 electrons in Q1 and Q2 respectively, as shown in \autoref{SEMs}b and c. This means we have effectively a single valence electron in each quantum dot (d-shell and p-shell, respectively) while the electrons in the inner shells stay inert during spin operations~\cite{leon2020coherent}. Evidence supporting the p- and d-shell structures is demonstrated in the Methods section. The particular choice of a p- and a d-shell electron is arbitrary, solely for a proof-of-principle. In an earlier study, we demonstrated the suitability of these shell configurations for single qubit operation, but a systematic study of the optimal number of electrons for a two-qubit system is out of the scope of our present work.

In general, EDSR control of qubits is heavily influenced by the details of the quantum dot confinement potential~\cite{camenzind2019spectroscopy}. 
We investigate these parameters performing an adiabatic inversion of the spins with a variable frequency microwave excitation, with an external magnetic field $B_0=\SI{1}{T}$. Firstly, the detuning $\varepsilon$ is varied across the (12,6)-(13,5) transition over a period of \SI{500}{\us}, such that a $\ket{\downarrow\downarrow}$ spin state is initialised adiabatically. We note that (12,6) provides a good initialisation because it is a spin-0 configuration, as confirmed by magnetospectroscopy (see supplemental material). Moreover, a large anticrossing gap between this (12,6) singlet and the $\ket{\downarrow\downarrow}$ state at (13,5) occupation is created by the difference in quantization axes between dots due to the micromagnet field gradient. We further improve the fidelity of this initialisation by simultaneously lowering $V_\mathrm{J}$, in order to enhance the energy gap between this target state and the (14,4) singlet. Subsequently, a chirped pulse of microwave excitation with variable frequency adiabatically flips one of the spins into an antiparallel configuration, creating either a $\ket{\downarrow\uparrow}$ or a $\ket{\uparrow\downarrow}$ state, if the frequency sweep matches the resonance frequency of the qubit. This spin flip is then read out by quickly changing $\varepsilon$ back to a (12,6) ground state, which will be blockaded by the Pauli principle unless the spin flip to the antiparallel configuration was successful.

\autoref{SEMs}d shows the nonlinear dependency of the qubit resonance frequencies with electric potentials (Stark shift).
Moreover, the efficiency of the adiabatic inversion of the spins depends on the intensity of the effective oscillatory field that drives Rabi oscillations. This is indicated by the colours in \autoref{SEMs}d, and shows that each qubit has a different optimal gate configuration, such that a sufficiently fast Rabi oscillation frequency is obtained to ensure good control fidelity. 
This dependence of the Rabi frequency on the gate voltage configurations was observed previously, and associated with the electron position shifting under the micromagnet field~\cite{leon2020coherent}. 
For more information on the method of choosing the optimal operation point, analysis of the Rabi efficiencies and coherence times of the qubits, refer to supplementary material.

\begin{figure*}
	\centering
	\includegraphics[width=\linewidth]{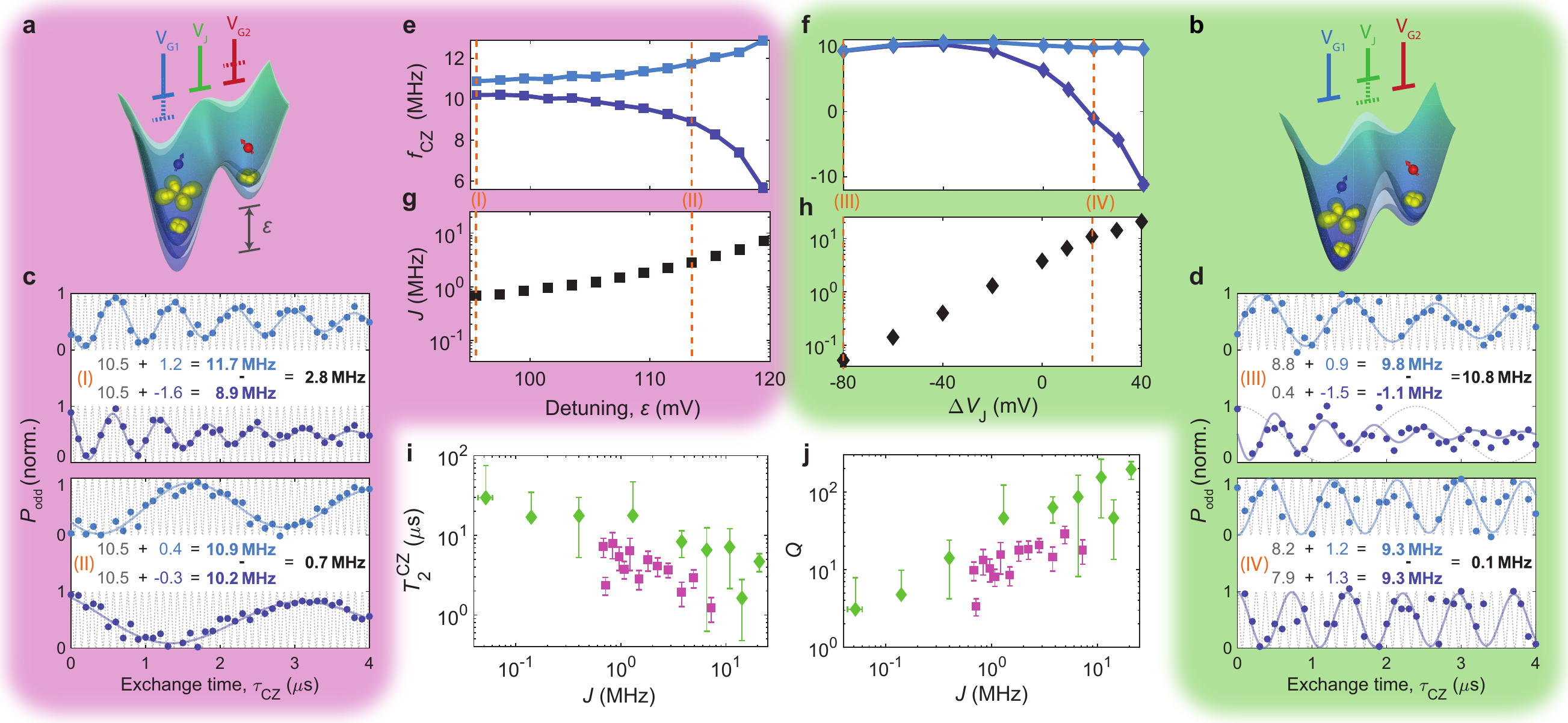}
	\caption{\textbf{Exchange control.} 
		\textbf{a,b}, Schematic showing the two different mechanisms to electrically control exchange coupling between quantum dots, by \textbf{(a)} voltage detuning between G1 and G2 gate, and \textbf{(b)} barrier control with J gate.
		\textbf{c,d}, Examples of CZ oscillations controlled via \textbf{(c)} detuning or \textbf{(d)} J gate. We apply a pulse sequence X1$-$CZ$-$X1, where X1 is a $\frac{\pi}{2}$ rotation around the $x$-axis of Q1, then measure the probability of an odd spin parity $P_\mathrm{odd}$. Green Roman numbers in each panel correspond to the applied voltage indicated in \textbf{(e-h)}, drawn as green dashed lines. Blue and red markers corresponds to the normalised measured $P_\mathrm{odd}$ with Q2 initialised as $\ket{\downarrow}$ or $\ket{\uparrow}$ respectively. The data is fitted using the equation $P_\mathrm{odd}= \frac{A}{2} (1-\cos(2\pi f_\mathrm{osc}t)e^{-t/T_2^\mathrm{CZ}})+b$. The Ramsey frequency $f_\mathrm{osc}$ is displayed as blue or red text on the panel. 
		In order to compensate the strong Stark shift induced by gate pulsing, we adopt different rotating frames, offset by a reference frequency $f_\mathrm{ref}$ between experiments, as presented in grey dashed curves behind each measurement data set. We extract the CZ frequency  $f_\mathrm{CZ} = f_\mathrm{ref}+f_\mathrm{osc}$ in a common frame and the difference between $f_\mathrm{CZ,Q2=\ket{\downarrow}}$ and $f_\mathrm{CZ,Q2=\ket{\uparrow}}$, which gives the exchange coupling frequency $J$, shown as black bold text.
		\textbf{e,f}, The oscillation frequency $f_\mathrm{CZ}$ as a function of \textbf{(e)} detuning or \textbf{(f)} J gate control. Blue and red line corresponds to Q1~=~$\ket{\downarrow}$ and $\ket{\uparrow}$, respectively.
		\textbf{g,h}, Extracted exchange oscillation frequency $J$.
		\textbf{i}, Damping time $T_2^\mathrm{CZ}$ of the measured oscillations as a function of exchange coupling $J$, for Q2~=~$\ket{\uparrow}$ and for detuning (yellow-green) and J gate control (purple).
		\textbf{j}, Quality factor $Q=J\times T_2^\mathrm{CZ}$ as a function of $J$, extracted from \textbf{(i)}.
	}\label{SpinControl}
\end{figure*}

The geometry of the MOS device studied here is known to lead to single electron wavefunctions that extend laterally approximately \SI{10}{\nm}~\cite{hensen2020silicon}, which is consistent with the large charging energy previously measured in this device when a second electron is added~\cite{leon2020coherent}. Since the nominal distance from the centre of G1 to the centre of G2 exceeds \SI{60}{\nm}, the inter-dot exchange coupling in the (1,1) charge configuration is predicted to be insufficient for quantum operations -- indeed, previous measurements in the same device reveal that exchange is only observed when the J gate is positive enough to form a dot under it~\cite{yang2020operation}. At the p- and d-shells, nonetheless, the Coulomb repulsion from the core electrons leads to a larger wavefunction for the valence electron. As a result, we are able to measure a sizeable interaction between distant qubits. The ability to control the inter-dot interaction is crucial for high fidelity two qubit gate operations~\cite{zajac2018resonantly}. High fidelity single qubit gates require low exchange coupling to ensure individual addressibility, while two qubit gates demand strong coupling for fast exchange oscillation with minimal exposure to noise. We explore two methods for controlling inter-dot interactions -- by detuning the quantum dot potentials~\cite{petta2005coherent,veldhorst2015twoqubit}, as shown in \autoref{SpinControl}a; or by directly controlling the inter-dot barrier potential via an exchange J gate~\cite{loss1998quantum,martins2016noise, zajac2018resonantly}, as in \autoref{SpinControl}b. 

\begin{figure*}
	\centering
	\includegraphics[width=\linewidth]{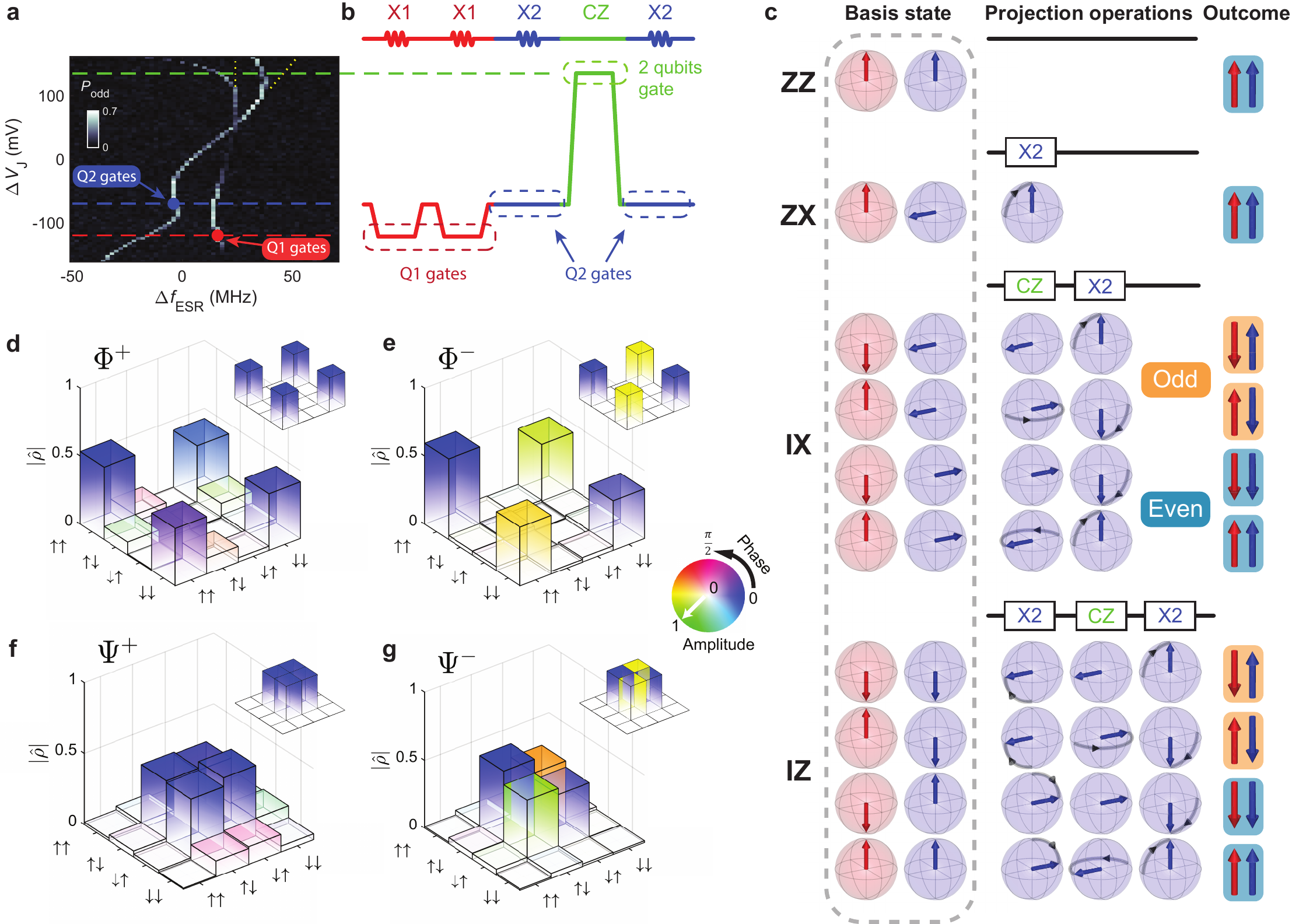}
	\caption{\textbf{Bell state tomography.} 
		\textbf{a}, Adiabatic inversion probability of both qubits as a function of detuned microwave frequency, where the carrier frequency is chosen to be the single qubit operation frequency for Q2, and J gate voltage $\Delta V_\mathrm{J}$, with qubits initialised in the $\ket{\downarrow\downarrow}$ state. Horizontal dashed lines represent J gate voltages applied for various single qubit and two qubit gates. Yellow dotted lines are a guide indicating the other resonance frequencies that would be observed at $\Delta V_\mathrm{J}>\SI{100}{\mV}$ if the spins were initialised randomly.
		\textbf{b}, Schematic of an example microwave and voltage pulse sequence for state tomography. It initialises the qubits as $\ket{\uparrow\downarrow}$ by performing two $\frac{\pi}{2}$ X1 pulses (all calibration is performed for $\frac{\pi}{2}$ pulses, such that a high fidelity $\pi$ pulse is obtained by composing it out of two $\frac{\pi}{2}$ gates, each starting and finishing at a common voltage $\Delta V_\mathrm{J}=\SI{-70}{\mV}$, which is shown as a blue dashed line in \textbf{(a)}), then perform IZ projection operation, by converting the parity readout into single qubit readout via a CNOT gate~\cite{yang2020operation}. Horizontal lines align with $\Delta V_\mathrm{J}$ from \textbf{(a)}.
		\textbf{c}, Example qubit states and operations required to obtain projections along the indicated axes. The first, two columns of Bloch spheres represent the eigenstates of Q1 (red) and Q2 (blue) before state tomography, while the rest illustrates the logic gate operations required for state tomography, before parity readout. For IX and IZ, all possible initial eigenstates are displayed, with parity results shown on the last column.
		\textbf{d-g}. Quantum state tomography of Bell states
		\textbf{(d)} $\Phi^+=\frac{\ket{\uparrow\uparrow}+\ket{\downarrow\downarrow}}{\sqrt{2}}$, \textbf{(e)} $\Phi^-=\frac{\ket{\uparrow\uparrow}-\ket{\downarrow\downarrow}}{\sqrt{2}}$, \textbf{(f)} $\Psi^+=\frac{\ket{\uparrow\downarrow}+\ket{\downarrow\uparrow}}{\sqrt{2}}$, \textbf{(g)} $\Psi^-=\frac{\ket{\uparrow\downarrow}-\ket{\downarrow\uparrow}}{\sqrt{2}}$.
		The height of the bars represents the absolute value of density matrix elements, while complex phase information is encoded in the colour map. Inset: bar graph of the ideal density matrix of the corresponding Bell state.
		The measured fidelities of each Bell state are \SIlist{87.1\pm2.8; 90.3\pm3.0; 90.3\pm2.4; 90.2\pm2.9}{\%}, from \textbf{(d)} to \textbf{(g)}, respectively.
	}\label{tomo}
\end{figure*}
	
For each method, the exchange intensity is measured by comparing the precession frequency of one qubit (target) depending on the state of the other qubit (control) with a Ramsey interferometry protocol. Due to the large difference in Larmor frequencies between quantum dots, only the $z$ components of the spins couple to each other, while the $x$ and $y$ components oscillate at different rates for each qubit and their coupling is on average vanishingly small~\cite{meunier2011efficient,thalineau2014interplay}. The measured oscillations shown in \autoref{SpinControl}c and d result from a combination of the exchange coupling and the Stark shift introduced by the gate pulses, measured with regard to a reference frequency $f_\mathrm{ref}$ which can be conveniently chosen to optimise the accuracy of our measurements (see supplementary material). The exchange coupling may be obtained by taking the difference between the resulting frequencies for the two states of the control qubit Q2 $\ket{\downarrow}$ and $\ket{\uparrow}$.

\autoref{SpinControl}e and f show the extracted oscillation frequencies as controlled by either the detuning $\varepsilon$ or the exchange gate voltage $V_\mathrm{J}$. The difference in oscillation frequencies corresponds to the exchange coupling and  can be tuned over two orders of magnitude, as seen in the extracted exchange coupling intensities in \autoref{SpinControl}g and h.We use this conditional control to implement the two-qubit CZ gate. The impact of exchange coupling on qubit coherence is quantified by extracting the decay time of the exchange oscillations $T_2^\mathrm{CZ}$, shown in \autoref{SpinControl}i as a function of the extracted exchange coupling for both CZ operation methods. We observe an improvement in the driven coherence times when the exchange control is performed by pulsing the J gate to control the inter-dot barrier, as compared to the detuning method. Since both methods can reach similar exchange frequencies, this results in an improvement in the quality factor of the exchange oscillations $Q=J\times T_2^\mathrm{CZ}$ as seen in \autoref{SpinControl}j, similarly to previously reported experiments~\cite{martins2016noise,reed2016reduced}. Throughout the rest of this work, we adopt the direct J gate-controlled exchange coupling method for the implementation of CZ logic gates.

As shown in \autoref{SEMs}d, both qubits possess a strongly non-linear Stark shift and large variation in the efficiency of the EDSR drive. Single qubit control fidelity in excess of \SI{99}{\%} was only achieved when the gate voltage configuration was tuned differently for each qubit, as indicated in the example gate sequence shown in \autoref{tomo}a. This leads to a major limitation -- single qubit gates must be performed sequentially, while the other qubit is left idling~\cite{culcer2009dephasing}, unable to be protected by refocusing techniques such as dynamical decoupling~\cite{meiboom1958modified,petta2005coherent} or pulse shaping~\cite{yang2019silicon}. Together with the two-qubit CZ gate, these gates span the two-qubit Clifford space (see \autoref{tomo}b for illustration). 

The strong Stark shift between operating points leads to a phase accumulation with regard to a reference frequency which must be accounted for in gate implementations (see supplementary material). In order to minimise the gate error introduced by resonance frequency shifts (due to electrical $1/f$ noise and $^{29}$Si nuclear spin flips), a number of feedback protocols are implemented. The following input parameters are monitored periodically and adjusted if necessary: SET bias voltage, readout voltage level, ESR frequencies of both qubits, phase accumulations at 5 different gate voltages for the logic gates, and exchange coupling. This results in a total of 10 feedback calibrations in each experiment. Further information on phase and exchange coupling feedback is provided in the supplementary section.

We gauge the quality of our gate set implementation by preparing Bell states and evaluating them through two-qubit state tomography~\cite{manko1997spin}. For a double quantum dot isolated from the reservoir, parity readout is used for the measurements~\cite{yang2020operation}, which implies that a readout step will contain the collective information of both qubits, or more precisely, the ZZ projection of the two qubits. In order to read out other projections, single and two qubit gate operations can be performed before readout. \autoref{tomo}c displays some key examples of such tomography protocols. The gate sequence illustrated in \autoref{tomo}b represents the example of an IZ measurement, which maps the spin state of the second qubit into the parities of the two-spin arrangement, regardless of the initial state of the first spin. In order to completely reconstruct the $4\times4$ density matrix of a two qubit system, 15 linearly independent tomography projections are required~\cite{rohling2013tomography} (the complete list is presented in the supplementary material). The results for each Bell state are shown in \autoref{tomo}d-g. The state preparation fidelities range from \SIrange{87.5}{90.3}{\%}, which compares favourably with state-of-the-art two spin qubit systems~\cite{zajac2018resonantly,watson2018programmable,huang2019fidelity}.

Our study highlights various advantages of multielectron qubits
which lead to efficient EDSR-based single qubit gates and extended reach of the exchange coupling between neighbouring qubits. The protocol for logic gates developed here leads to promising fidelities for Bell state preparation, but its use in longer computations would be impacted by the inability to refocus the spin that is not being manipulated. This problem can be solved by designing a more efficient EDSR strategy without the need to optimise the gate configuration, or by using an antenna to produce microwave magnetic field-based electron spin resonance~\cite{veldhorst_addressable_2014}. 
The ability of additional core electrons to screen charge disorder at the Si/SiO$_2$ interface~\cite{barnes2011screening,hu2001spinbased}, as demonstrated here, indicates that multielectron qubits offer a promising pathway for near term demonstrations of quantum processing in silicon.

\section{References}
\bibliographystyle{naturemag}
\bibliography{13e5e_v2}

\clearpage

\section{Methods}
\subsection{Magnetospectroscopy of an isolated double quantum dot}

\begin{extfig*}
	\centering
	\includegraphics[width=\linewidth]{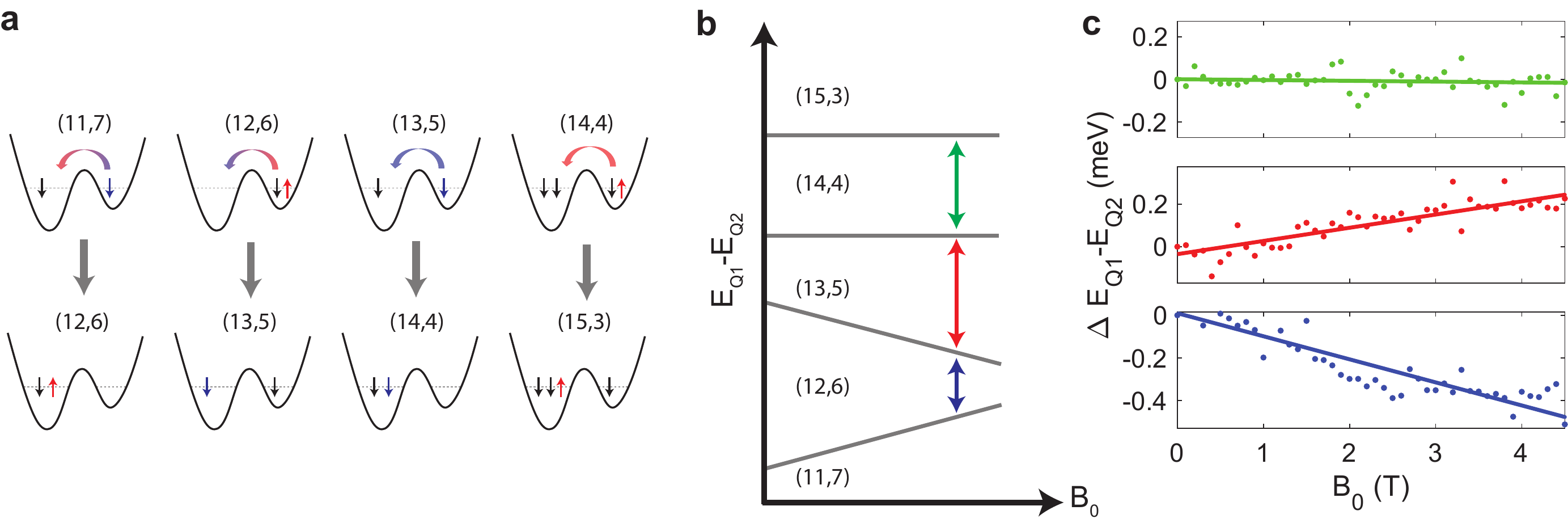}
	\caption{\textbf{Magnetospectroscopy of an isolated double quantum dot.} 
		\textbf{a}, Estimated spin state of the active valence electrons before (top row of schematics) and after (bottom row of schematics) an inter-dot charge transition at corresponding electron number ($m$,$n$), where $m$ and $n$ represent the total number of electrons in Q1 and Q2 respectively. Coloured arrows represents the electron which participates in charge transition, with blue and red indicate spin down and up, respectively. 
		\textbf{b}, Illustration of energy difference between Q1 and Q2 as a function of applied magnetic field $B_0$, as corresponding electron numbers in each dot. 
		\textbf{c}, Extracted experimental magnetospectroscopy data, with each colour corresponding to the energy difference in charge transition shown in \textbf{(b)}.
	}\label{supp-magspec}
\end{extfig*}

From the single dot shell structure~\cite{leon2020coherent}, one can try to predict which double dot occupations will lead to a single spin \textonehalf\ qubit in each dot. But in order to confirm that the spin structure of the double dot can be extrapolated from single dot results, we obtain the spin ordering of the dots performing magnetospectroscopy. Traditionally, magnetospectroscopy is performed studying the shifts of chemical potentials of each dot as a function of the externally applied magnetic field. This assumes that the quantum dot is in diffusive equilibrium with a reservoir (same chemical potential). Such reservoir is assumed to be spinless, such that its chemical potential does not shift with magnetic field and the absolute shift in dot chemical potential with magnetic field can be assessed. In our system, the two dots are in equilibrium with each other, but all transitions conserve the total number of electrons in the double dot system (isolated double dot) -- there is no reference reservoir, as shown in \autoref{supp-magspec}a. Therefore, only relative Zeeman shifts are observed.

The hypothetical field dependencies, assuming that the shell structure from Ref.~\onlinecite{leon2020coherent} holds, are shown in the energy diagram in \autoref{supp-magspec}b. The measured magnetospectroscopy results in \autoref{supp-magspec}c confirm our assumption. In particular, the (13,5) charge configuration consists indeed of single spin-\textonehalf\ states in both dots, each atop an inert closed shell of spin 0.

Note that the leverarm we extracted from the slope in \autoref{supp-magspec}c is the sum of leverarm from Q1 and Q2, approximately $\alpha_\mathrm{Q1}+\alpha_\mathrm{Q2} = \SI{0.53}{\eV\per\V}$. Differences in leverarm $\alpha_\mathrm{Q1}-\alpha_\mathrm{Q2}$ cannot be obtained from this method. 

\subsection{Adiabatic inversion and qubit operation points}

\begin{extfig*}
	\centering
	\includegraphics[width=0.9\linewidth]{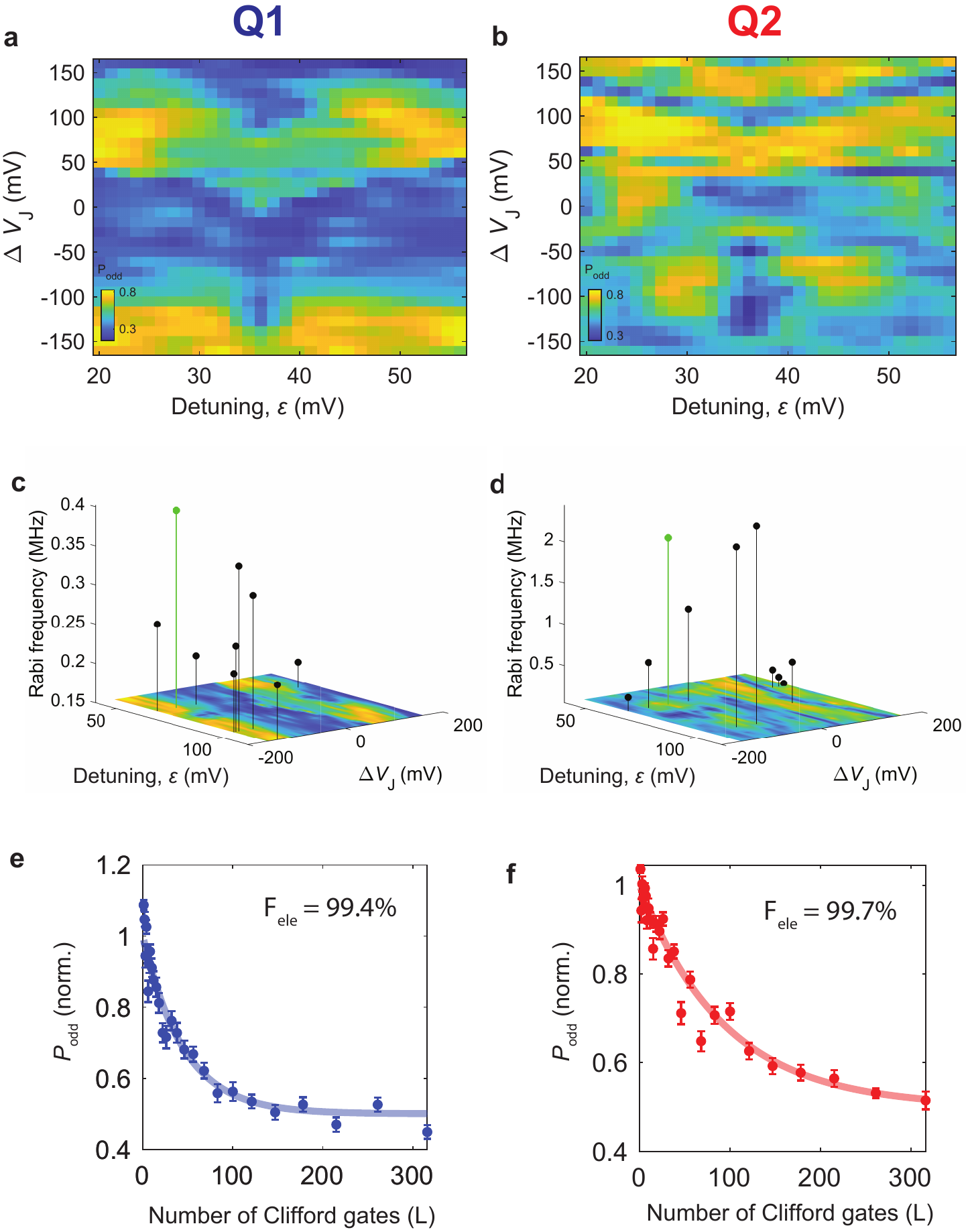}
	\caption{\textbf{Single qubit operation voltage.} 
		\textbf{a,b}, Adiabatic inversion probability of \textbf{(a)} Q1 or \textbf{(a)} Q2 as a function of detuning and J gate voltage, with interpolation.
		\textbf{c,d}, Rabi frequency $f_\mathrm{Rabi}$ for selected detuning and J gate voltage combinations, with the 2D plot of panels \textbf{(a)} and \textbf{(b)} copied at bottom of $x$-$y$ plane.
		\textbf{e,f}, Single qubit randomised benchmarking for \textbf{(e)} Q1 or \textbf{(f)} Q2 at voltages at the green marker in panels \textbf{(c)} and \textbf{(d)}, respectively. 
	}\label{supp-oppt}
\end{extfig*}

In order to achieve single qubit EDSR control fidelities exceeding \SI{99}{\%}, compliant with the demands for quantum error correction in the surface code architecture, we must adjust the inter-dot detuning and J gate voltage such that we achieve the most efficient Rabi drive for both Q1 and Q2. 

We perform an adiabatic spin inversion experiment by sweeping the microwave frequency applied to the EDSR gate electrode (in our case the Co magnet) at fixed power, such that when each of the qubit resonance frequencies $f_\mathrm{ESR}$ is found, that spin is flipped with an efficiency given by the comparison between the sweeping speed and the Rabi frequency (limited by the spin relaxation time)~\cite{laucht2014high}. This is observed as an increase in the probability of measuring an odd parity readout after preparing the even initial state $\ket{\downarrow\downarrow}$, with an example shown in \autoref{tomo}a. This permits us to determine the resonance frequencies, as well as the region of high qubit fidelity, as a function of detuning and J gate voltage.

The colour scale in \autoref{SEMs}d shows the extracted adiabatic inversion probability of each qubit at various detuning and J gate voltages. We interpolated these probabilities and plotted them again in \autoref{supp-oppt}a and b.
At first glance, we notice that $P_\mathrm{odd}$ is symmetric along the axis of detuning $\varepsilon=\SI{37}{\mV}$, implying that detuning the dots in either direction has the same effect on dot shape and spin behaviour.

The strategy to quickly calibrate the ideal operation points is to choose a few potential operation points on the 2D map where $P_\mathrm{odd}$ shows a high adiabatic inversion probability, and measure the Rabi oscillation frequency at a fixed microwave power. We then choose the highest Rabi frequency point that meets some constrains. Firstly, for individual addressability by frequency modulation, the ESR frequency $f_\mathrm{ESR}$ of both qubits should be at least \SI{10}{\MHz} apart, which means $\Delta V_\mathrm{J} <\SI{20}{\mV}$ or $>\SI{100}{\mV}$ in \autoref{tomo}a. Also, we would like to minimise the exchange coupling during single qubit operation, which is achieved for $\Delta V_\mathrm{J} <\SI{-20}{\mV}$, setting $J<\SI{1}{\MHz}$ as observed from \autoref{SpinControl}h. As a result, we are generally limited to the bottom half of the 2D map in \autoref{supp-oppt}a and b. Ideally, we would like to choose an optimal operation point such that we can perform single qubit operation on both Q1 and Q2 (see main text for detail). However, there is no observable voltage range from \autoref{supp-oppt}a and b where both qubits gives high $P_\mathrm{odd}$ under the constrains mentioned above.

A few detuning and J gate voltage combinations with $P_\mathrm{odd}>0.42$ are chosen for each qubit, and Rabi frequencies are extracted in \autoref{supp-oppt}c and d. The green markers from the plots are the operation points chosen for single qubit randomised benchmarking, with results presented in \autoref{supp-oppt}e and f. Qubits Q1 and Q2 have control fidelities $F_\mathrm{Q1}=\SI{99.4\pm0.17}{\%}$ and $F_\mathrm{Q2}=\SI{99.7\pm0.10}{\%}$, respectively. Note that the operation point chosen for Q2 is not the one with the absolute maximum Rabi frequency, as we also would like to minimise gate voltage fluctuation when ramping between Q1 and Q2 logic gate operations. We observe a significant influence of ramping range on the final outcome of the Bell state preparation, but a thorough evaluation of this source of error is not warranted, since this relates to instrument limitations.

Coherence times $T_2^*$ for Q1 and Q2 at the chosen operation points are \SIlist{13.7\pm2.0;8.4\pm3.3}{\us}, respectively, while $T_2^\mathrm{Hahn}$ are \SIlist{50.0\pm15.2;94.6\pm18.7}{\us}, respectively. 

\subsection{Exchange oscillation, coherence and Q factors of interacting spins}

The oscillations observed from Ramsey-like experiments in the main text \autoref{SpinControl}c, d are due to difference in precession frequency of the qubits in the period between $\frac{\pi}{2}$-pulses. The difference in frequencies arises from both Stark shift, which is in the order of \SI{10}{\MHz} in our experiments, and exchange coupling $J$, between \SI{100}{\kHz} and \SI{10}{\MHz}.
As a result, the total Ramsey frequency will be dominated by Stark shift, making the $J$-coupling effect difficult to observe without a high resolution scan of precession time.
Therefore, we adjust the phase of the second $\frac{\pi}{2}$-pulse to match a rotating frame of reference which is not the same as the qubit Q1 precession frequency $f_\mathrm{Q1}$, but instead it is offset by a value $f_\mathrm{ref}$ chosen to reduce the impact of the Stark shift to the oscillation observed in experiment. This reference frequency is adjusted \textit{ad hoc} between different experiments in order to facilitate the extraction of the exchange coupling effect.

In the left panel of \autoref{SpinControl}, where the quantum dots are detuned, $f_\mathrm{ref}$ is set to \SI{10.5}{\MHz} throughout the experiment. 
However, for direct J gate controlled CZ, the oscillation frequency varies across a range of \SI{20}{\MHz}, as shown in \autoref{SpinControl}f. In order to capture the oscillation data efficiently, we assign various $f_\mathrm{ref}$ for each $\Delta V_\mathrm{J}$ targeting a shift of approximately \SI{-1}{\MHz} from the CZ frequency $f_\mathrm{CZ}$ (which could differ depending on whether the control spin is up or down).

In a qubit rotating frame, positive and negative phase accumulation will result in the same Ramsey oscillation if only a single measurement projection is taken. To determine the sign of ESR frequency shift, we repeat every Ramsey experiment with additional phase shift on the second $\frac{\pi}{2}$ pulse, in order to extract X,$-$X,Y,$-$Y projections of the qubit. Note that all four measurements are taken in a interleaved fashion to minimise the impact of quasi-static noise.

\subsection{Measurement feedback} 
\label{s:feedback}

Low frequency noise is a major limitation for high fidelity operation of qubits in MOS devices~\cite{huang2019fidelity}. An efficient approach to mitigate high amplitude noise that occurs in a sub-\si{\Hz} scale is to recalibrate the most critical qubit control parameters periodically.

There are 10 parameters that require feedback throughout the experiments due to the intricate way by which the qubit operations are defined with different gate configurations targeting the optimisation of each qubit. These parameters are the SET Coulomb peak alignment, the readout level set by the dot gate, both qubit ESR frequencies, a total of five relative phases acquired when pulsing between operating points, and the exchange coupling controlled by the J gate. The SET feedback is used to maintain its high sensitivity during charge transition, while read level feedback is to ensure the readout is done within a Pauli spin blockade region for parity readout. SET and readout level feedbacks are performed with first order corrections, with a predefined target SET current. SET top gate voltage $V_\mathrm{ST}$ and read level voltage (controlled via $V_\mathrm{G1}$) are updated based upon the difference between measured current and target current.

We adopt the ESR frequency tracking protocol from Ref.~\onlinecite{huang2019fidelity} in order to follow the resonance frequency jumps due to quasi-static noise such as hyperfine coupling with residual $^{29}$Si nuclear spin in the silicon wafer, as well as low frequency electrical noise. We perform checks of each of the two resonance frequencies shown in \autoref{tomo}a independently every 10 measurement data points. If the spin rotation is unsuccessful at the assumed resonance frequency, we recalibrate the frequency with a series of Ramsey experiments.

In \autoref{tomo}a, the ESR frequency shift $\Delta f_\mathrm{ESR}$ is taken as \SI{0}{\MHz} at the microwave driving frequency that matches the resonance frequency of Q2 at voltage $\Delta V_\mathrm{J}=\SI{-70}{\mV}$, which is the operating point for Q2. At all the other operation points where $\Delta f_\mathrm{ESR}$ is non-zero, a phase will accumulate due to variations in precession frequency. Since our Clifford set requires 3 operation voltages, each with two phases for Q1 and Q2 to track, excluding the reference frequency $f_\mathrm{ESR}=f_\mathrm{Q2}$, that results in 5 phase accumulations to recalibrate.

Although phase accumulation can be calculated by the extracted ESR frequency ($\Delta f_\mathrm{ESR}$) and gate time $t_\mathrm{g}$, i.e. $\phi = \Delta f_\mathrm{ESR}\times t_\mathrm{g}$, such method assumes an instantaneous step from one gate voltage to another, which in reality is limited by the \SI{80}{\MHz} bandwidth of the measurement cable, meaning during the ramp both qubits spend a non-negligible amount of time in an intermediate voltage state, accumulating phases that are non-trivial to calculate, especially when the Stark shift is highly non-linear as seen in \autoref{tomo}a. Moreover, it is unclear whether the low frequency noise will affect the overall shape of the gate dependency of the resonant frequencies.

In quantum computing, all operations can be performed by a sequence of gates taken from a primitive gate set. The processing unit is fully calibrated if all the primitive gates are calibrated individually. \autoref{table:phasecali} shows the pulse sequences required to extract each of the 5 phases accumulated, each associate with certain qubit and primitive gates. 

\begin{table*}
\centering
\begin{tabular}{ |c| c !{\vrule width 2pt} c |c|| c | c ||c|} 
\hline
Level & $\Delta V_\mathrm{J}$ (\si{\mV}) & Q1 & target gate & Q2 & target gate & Primitive gate \\ \hline
1 & -120 & X1$^2$ & I1 & X2$-$I1$-$X2 & I1 & X1,Y1 \\ \hline
2 & -70 & X1$-$I2$-$X1 & I2 & N/A & N/A & X2,Y2\\ \hline
3 & 130 & X1$-$CZ$-$X1 & CZ & X2$-$CZ$-$X2 & CZ & CZ\\\hline
\end{tabular}
\caption {\textbf{Pulse sequences for qubit phase calibration.}
Pulse sequences used to extract phase accumulation while idling. $\Delta V_\mathrm{J}$ (\si{\mV}) is referenced from \autoref{tomo}a. Element at column Q$n$ row $\Delta V_\mathrm{J}$ corresponds to pulse sequence required to extract phase accumulated in qubit $n$ when inter-dot barrier gate voltage is at $\Delta V_\mathrm{J}$. $Rn$ represents a $\frac{\pi}{2}$ rotation around $R$-axis on qubit $n$, with $R\in\{\mathrm{X},\mathrm{Y}\}$, while I$n$ means identity gate with $\Delta V_\mathrm{J}$ equals to the voltage where single qubit operation is performed for qubit $n$. }
\label{table:phasecali}
\end{table*}

Phase calibration is performed every ten measurements, after the ESR frequencies are updated. In each calibration, the corresponding pulse sequence from \autoref{table:phasecali} is applied with various phases $\phi$ for the last $\frac{\pi}{2}$ pulse with respect to the other pulses. The results are then fitted with a function $P_\mathrm{odd} = A\cos(2\pi(\phi-\phi'))+b$, where $A$ and $b$ are fitting constants related to the oscillation visibility and dark counts, while $\phi'$ is the phase accumulated from the target gate. Since this protocol may rely on multiple primitive gates in a sequence, the phase associated with each gate in \autoref{table:phasecali} has to be calibrated following a certain order , to ensure the phase extracted corresponds to one particular primitive gate only.
These phases $\phi'$ will be used for compensation of unwanted accumulated phases as we apply the corresponding Clifford gates in the experiment.

\subsection{Exchange coupling feedback}

\begin{extfig*}
	\centering
	\includegraphics[width=\linewidth]{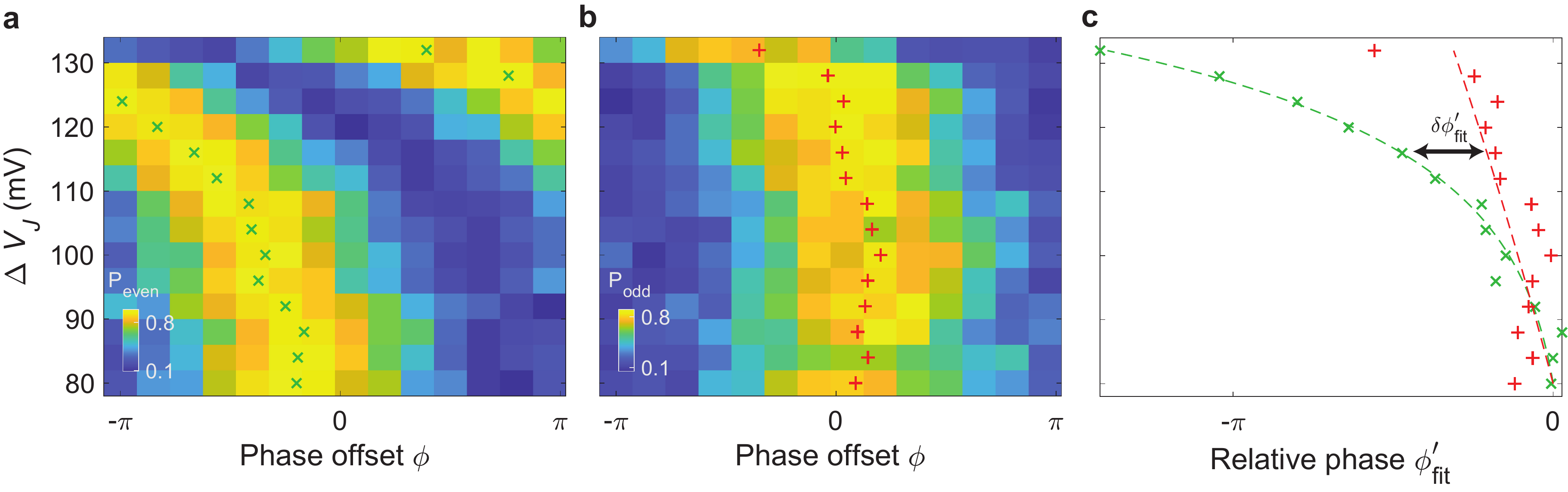}
	\caption{\textbf{Phase accumulation from CZ operation.} 
		\textbf{a,b}, Parity readout probability as a function of exchange gate voltage $\Delta V_\mathrm{J}$ and phase offset $\phi$, for duration of $\tau_\mathrm{CZ}=\SI{160}{\ns}$, with gates \textbf{(a)} X1$-$CZ$-$X1$(\phi)$ or \textbf{(b)} X2$^2-$X1$-$CZ$-$X1$(\phi)$ applied. $\phi$ represent the phase offset of the second X1 pulse with respect to the first within the same sequence. Marker in each row indicate the fitted phase $\phi'$ from a $P_\mathrm{even/odd} = A\cos(2\pi(\phi-\phi'))+b$, where $A$,$b$ and $\phi'$ are constants. Note that since the control qubit is initilised into opposite spin state and parity readout is used, opposite parity is extracted for the two cases in order to obtain the same single spin information from the target qubit. 
		\textbf{c}, Fitted phase $\phi'$ from \textbf{(a)} (green `$\times$') and \textbf{(b)} (red `$+$'), which are fitted with equation $\phi_\mathrm{fit}'=A\exp(-b \Delta V_\mathrm{J})+c$, where $A$, $b$ and $c$ are constants. Both graphs are offset to zero phase at $\Delta V_\mathrm{J}=\SI{80}{\mV}$.
	}\label{Jfeedback}
\end{extfig*}

The exchange coupling $J$ may fluctuate between experiments due to low frequency electrical noise, which can be compensated by monitoring and recalibrating the CZ gate operation with a feedback protocol. The sampling rate of the arbitrary waveform generator (AWG) and microwave IQ modulation used here, \SIlist{8;10}{\ns} respectively, limit our gate operation times to the least common multiple of these two, $\tau_\mathrm{CZ}=\SI{40}{\ns}$, or any multiples of that. This means that updating the CZ exchange time $\tau_\mathrm{CZ}$ is not accurate enough for high fidelity operation. Instead, we update the inter-dot barrier gate voltage $V_\mathrm{J}$, which compensates the change in $J$ while leaving $\tau_\mathrm{CZ}$ unchanged.

The initial calibration method is as follows: two CZ identical sequences are performed, each one with an opposite control qubit state (spin down or up). We vary the readout projection angles $\phi$ and fit the parity readout probability to a sinusoidal wave similar to the case of the phase feedback, which we use to extract the phase offset $\phi_\mathrm{fit}'$. The difference in phase accumulated in the control spin down and up cases are due to the composition of an exchange coupling from the CZ operation and from the extra X2$^2$ gate necessary for the control spin up calibration step. The latter can be compensated by re-scaling $\phi_\mathrm{fit}'$ to 0 at low exchange coupling regime.

This experiment is repeated with various exchange gate voltages $\Delta V_\mathrm{J}$, as shown in \autoref{Jfeedback}a and b, while the resulting phases, are plotted on \autoref{Jfeedback}c, along with an exponential fit. The difference between the two lines in \autoref{Jfeedback}c are the phase contributed from exchange coupling $J$, which can be calculated from $J=\frac{\var\phi_\mathrm{fit}'}{\tau_\mathrm{CZ}}$.

Upon choosing the desired value of $J$ with the associated $\Delta V_\mathrm{J}$, which should correspond to a $\var\phi^\prime_\mathrm{fit}=\pi$ phase difference between the two initial states, a feedback protocol can be implemented to recalibrate $J$ periodically. The feedback protocol is similar to the initial calibration mentioned above, but optimised for speed by focusing on a smaller range of $\Delta V_\mathrm{J}$, and the exponential fit used in  \autoref{Jfeedback}c is replaced with a linear fit. With that, the value of $\Delta V_\mathrm{J}$ is updated using the fit in order to maintain the same exchange coupling strength $J$.

This exchange coupling feedback is performed after ten measurements, immediately after the phase calibration step. Note that the pulse sequence used in \autoref{Jfeedback}a is identical to the one in \autoref{table:phasecali}. Therefore, the X1$-$CZ$-$X1 sequence is omitted from the phase calibration stage, but extracted from the subsequent exchange coupling feedback stage.

\begin{extfig*}
	\centering
	\includegraphics[width=0.8\linewidth]{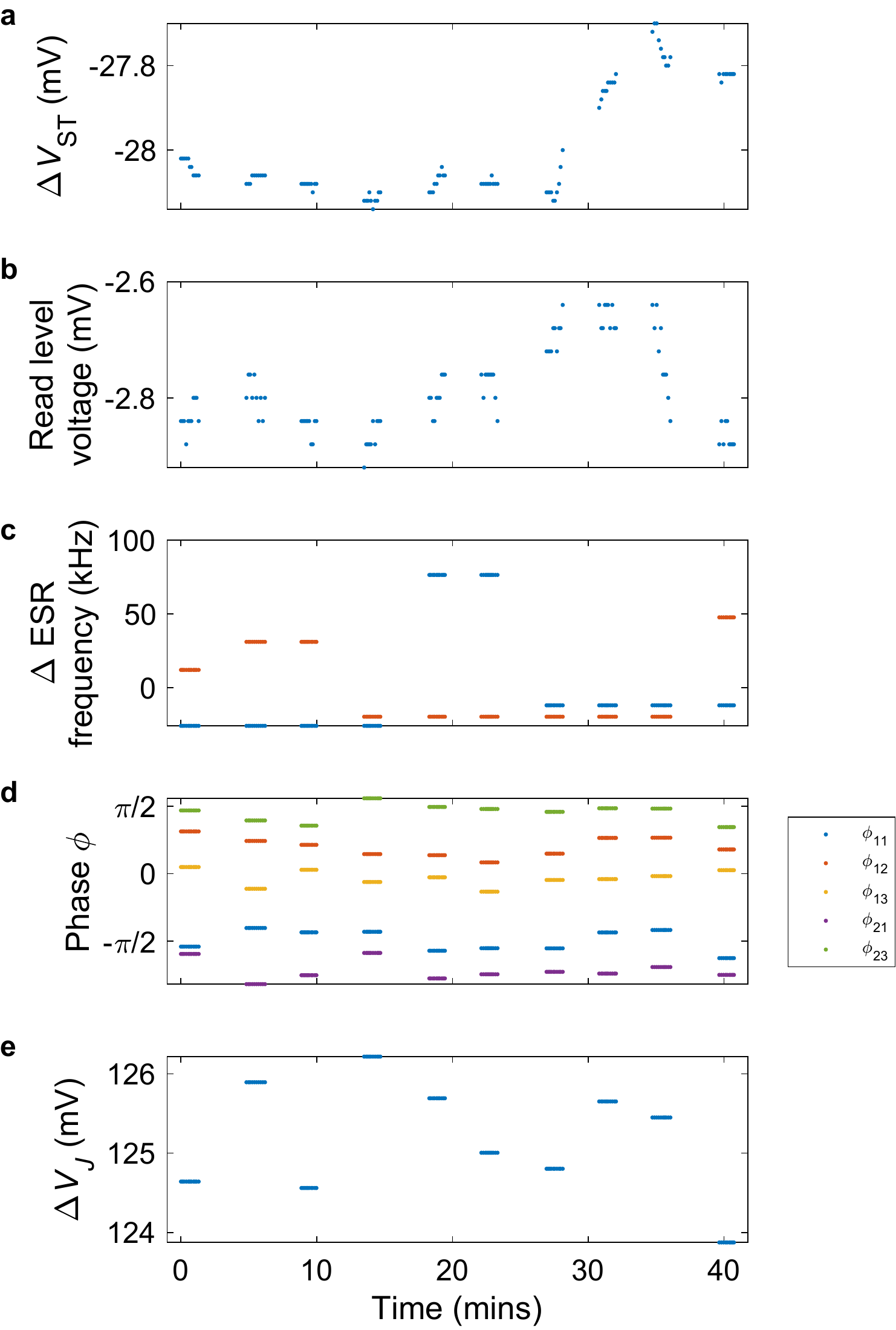}
	\caption{\textbf{Parameters tracking over measurement time.} 
		Various parameters are recorded while Bell state tomography of \autoref{tomo}g is running.
		\textbf{a}, ST gate voltage $\Delta V_\mathrm{ST}$ from SET feedback.
		\textbf{b}, G1 gate voltage $\Delta V_\mathrm{G1}$ during readout, from parity readout feedback.
		\textbf{c}, Change in resonance frequency $\Delta_\mathrm{ESR}$ for Q1 (blue) and Q2 (red).
		\textbf{d}, Phase accumulation $\phi_{MN}$ from the target gate in \autoref{table:phasecali}, where $M$ and $N$ represents the level and qubit in the table, respectively.
		\textbf{e}, J gate voltage $\Delta V_\mathrm{J}$ required to maintain a phase difference of $\var\phi'_\mathrm{fit}=\pi$ at $\tau_\mathrm{CZ}=\SI{160}{\ns}$ during CZ operation.
	}\label{feedbackvstime}
\end{extfig*}

\autoref{feedbackvstime} is an example of a Bell state tomography experiment, with all ten feedback loops active, and the variation of the respective parameters over 40 minutes of laboratory time. The parameters that are calibrated only every ten measurements have larger gaps between data points.

\subsection{Two qubit tomography with parity readout}

A two qubit density matrix is a $4\times4$ matrix spanning a $4^2-1=15$ dimensional space and requires 15 linearly independent projection measurements. Ref.~\onlinecite{seedhouse2020parity} gives a detailed explanation on how to perform two-qubit state tomography using parity readout. \autoref{table:15proj} lists the gate operation sequences adopted here for each of the 15 projection measurements, using a combination of primitive gates described in the main text.
\begin{table}
\centering
\begin{tabular}{ | c | c |} 
\hline
Projection & Operations \\
\hline
ZZ & I\\
\hline
YZ & X1\\ \hline
XZ & Y1\\\hline
ZY & X2\\\hline
ZX & Y2\\\hline
YY & X1$-$X2\\\hline
YX & X1$-$Y2\\\hline
XY & Y1$-$X2\\\hline
XX & Y1$-$Y2\\\hline

YI & CZ$-$X1\\\hline
XI & CZ$-$Y1\\\hline
IY & CZ$-$X2\\\hline
IX & CZ$-$Y2\\\hline
ZI & X1$-$CZ$-$X1\\\hline
IZ & X2$-$CZ$-$X2  \\ 
\hline
\end{tabular}\\
\caption {\textbf{Gate operations for parity readout.}
List of operations required for a complete state tomography via parity readout, with each row representing the projection axis of interest for a two-qubit system, and the sequence of gate operations required prior to readout.}
\label{table:15proj}
\end{table}

\subsection{Fidelity estimation}

In order to accurately estimate the fidelity of the control steps in preparing a Bell state, some post-processing techniques are applied to the outcome of the measured odd parity probability $P_\mathrm{odd}$ corresponding to the 15 projections from \autoref{table:15proj}. 

Firstly, we factor in the errors associated with state initialisation and measurement (SPAM error), by renormalising the parity readout probability of the two qubits for ZZ readout.

Next, we reconstruct the density matrix from the measurement data. Let $E_\upsilon$ be the measurement outcome projector, $\rho$ be density matrix, $p_\upsilon$ be the measurement probability, where $\upsilon=1...30$ (notice that measurements of the projector $P_{MN}$, where $M,N\in \{\mathrm{I,X,Y,Z}\}$, produce not only probability $p_{MN}$, but also $p_{-MN} = 1-p_{MN}$, so that 15 projections yield 30 probabilities). We define a matrix $A$ as
\begin{equation}
    A = \mqty(
    \vec{E}_1^\dagger \\
    \vec{E}_2^\dagger \\
    \vdots\\
    \vec{E}_{30}^\dagger,
    )
\end{equation}
where $\vec{E}_\upsilon^\dagger$ stands for the vectorised form of the projection $E_\upsilon$.

Similarly, all elements of $\rho$ can also be vectorised.
This yields the relation:
\begin{equation}
\begin{split}
    A\vec{\rho} &= 
    \mqty(
    \vec{E}_1^\dagger \vec{\rho}  \\
    \vec{E}_2^\dagger \vec{\rho} \\
    \vdots\\
    \vec{E}_{30}^\dagger \vec{\rho} 
    )
    =
    \mqty(
    \tr\{ E_1^\dagger \rho \} \\
    \tr\{ E_2^\dagger \rho \} \\
    \vdots\\
    \tr\{ E_{30}^\dagger \rho \}
    )
    =
    \mqty(
    P( E_1 |\rho ) \\
    P( E_2 |\rho )  \\
    \vdots\\
    P( E_{30} |\rho ) 
    ) \\
    &\approx
    \mqty(
    p_1\\
    p_2  \\
    \vdots\\
    p_{30}
    )
    = \vec{p}
    \end{split}
\end{equation}
With matrix $A$ constructed from our choice of measurement projection, and $\vec{p}$ from measurement data. We then perform a (pseudo) linear inversion to estimate the density matrix $\hat{\rho}$.

Since the matrix computed numerically by linear inversion can be an unphysical state for a qubit (leading to a measured matrix $\vec{p}$ that does not have the properties of a density matrix), a maximum likelihood technique is used to numerically estimate the density matrix \citemethod{altepeter2004qubit} under several constrains. A legitimate qubit density matrix must be non-negative definite, have a trace of one and be Hermitian. These conditions are met if we write the density matrix as~\citemethod{altepeter2004qubit}:
\begin{equation}
   \hat{\rho} = \frac{T^\dagger T}{\tr\{T^\dagger T \}}
\end{equation}
where
\begin{equation}
   T = \mqty(
t_1 & 0 & 0 & 0\\
t_5+it_6 & t_2 & 0 & 0 \\
t_{11}+it_{12} & t_7+it_8 & t_3 & 0 \\
t_{15}+it_{16} & t_{13}+it_{14} & t_9+it_{10} & t_4
)
\end{equation}
and $t_1..t_{16}$ are real numbers. To find these values, we apply a maximum likelihood estimation, with the cost function
\begin{equation}
    L(t_1,t_2,...t_{16}) = \sum_\upsilon \frac{(\expval{\hat{\rho}(t_1,t_2,...t_{16})}{\psi_\upsilon}-n_\upsilon)^2}{2\expval{\hat{\rho}(t_1,t_2,...t_{16})}{\psi_\upsilon}}
\end{equation}
where $\psi_\upsilon$ is the vectorised measurement matrix with $\upsilon = 1...30$ and $n_\upsilon$ are the measurement probabilities. We start our search inputing the density matrix resulting from the pseudo-linear inversion described before and proceed to numerically optimise $L$ as a function of $t_1,t_2,...t_{16}$. The resulting elements will give our final density matrix.

The fidelity of a Bell state is calculated then from the definition $F(\rho,\hat{\rho}) = (\tr\{\sqrt{\sqrt{\rho}\hat{\rho}\sqrt{\rho}} \})^2$, where $\rho$ and $\hat{\rho}$ are the ideal and measured density matrices, respectively.

\section{Data availability}
The data that support the findings of this study are available from the authors on reasonable request, see author contributions for specific data sets.
	
\section{Code availability}
The code that support the findings of this study are available from the authors on reasonable request, see author contributions for specific code sets.

\section{References}
\bibliographystylemethod{naturemag}
\bibliographymethod{13e5e_v2}

	\section{Acknowledgments}
	We acknowledge support from the Australian Research Council (FL190100167 and CE170100012), the US Army Research Office (W911NF-17-1-0198), Silicon Quantum Computing Pty Ltd, and the NSW Node of the Australian National Fabrication Facility. The views and conclusions contained in this document are those of the authors and should not be interpreted as representing the official policies, either expressed or implied, of the Army Research Office or the U.S. Government. The U.S. Government is authorized to reproduce and distribute reprints for Government purposes notwithstanding any copyright notation herein.
		J. C. and M. P. acknowledge support from the Canada First Research Excellence Fund and in part by the National Science Engineering Research Council of Canada.

	\section{Author Contributions}
	
	R.C.C.L. and C.H.Y. performed the experiments.
	J.C.L., R.C.C.L., J.C.C.H., C.H.Y. and M.P.\nobreakdash-L. designed the micromagnet, which was then simulated by J.C.L and M.P.\nobreakdash-L.
	J.C.C.H. and F.E.H. fabricated the device with A.S.D's supervision.  
    K.M.I. prepared and supplied the $^{28}$Si epilayer. 
    J.C.C.H., W.H. and T.T. contributed to the preparation of experiments.
    R.C.C.L., C.H.Y., A.S. and A.S.D. designed the experiments, with J.C.L., M.P.\nobreakdash-L., W.H., T.T., A.M. and A.L. contributing to results discussion and interpretation. 
    R.C.C.L., A.S., and A.S.D. wrote the manuscript with input from all co\nobreakdash-authors.
    R.C.C.L. and J.Y.H. contributed in device visualisation in the manuscript.

\end{document}